\documentclass[letterpaper,11pt]{article}
\usepackage{amsmath,amssymb}
\usepackage{bm}
\usepackage{graphicx,epsfig,psfrag}
\usepackage{cite}

\makeatletter
 
 \@addtoreset{equation}{section}
\makeatother

\newcommand{\gsim}{\mathrel{\lower2.5pt\vbox{\lineskip=0pt\baselineskip=0pt
                   \hbox{$>$}\hbox{$\sim$}}}}
\newcommand{\lsim}{\mathrel{\lower2.5pt\vbox{\lineskip=0pt\baselineskip=0pt
                   \hbox{$<$}\hbox{$\sim$}}}}

\newcommand{\sla}[1]{{\raise.15ex\hbox{$/$}\kern-.57em #1}}
\newcommand{\Sla}[1]{\kern0.12em{\raise.15ex\hbox{$/$}\kern-.74em #1}}

\newcommand{\beq}{\begin{eqnarray}}
\newcommand{\eeq}{\end{eqnarray}}

\newcommand{\me}{\mbox{${\rm \not\! E}$}}

\begin{document}


\begin{titlepage}

\setcounter{page}{0}

\begin{flushright}
EDINBURGH-2009/03
\end{flushright}

\vskip 3cm

\begin{center}

{\huge\bf Discovering a Nonstandard Higgs in a Standard Way}

\vskip 1cm

{\large {\bf Spencer Chang} $^a$ and {\bf Thomas Gr\'egoire} $^b$}

\vskip 0.5cm

$^a${\it Department of Physics, University of California at Davis, CA 95616\\ and Center for Cosmology and Particle Physics, New York University, New York, NY 10003}\\[.2cm]
$^b${\it SUPA, School of Physics, University of Edinburgh, \\  Edinburgh, EH9 3JZ, Scotland, UK}\\

\abstract{
In this note, we explore the Tevatron and LHC detection prospects for a nonstandard Higgs whose decays contain both missing and visible energy. In particular, we consider the decay $h \rightarrow \chi_2 \chi_1$ where $\chi_1$ is stable and $\chi_2$ decay back to $\chi_1$ and additional visible particles.  We focus on adapting standard searches for both Higgses and supersymmetry to these decays.  In these channels, the Tevatron has little reach, but the LHC has good prospects for decays that involve leptons or neutrinos.  In particular, the standard WW* search for a Higgs produced in vector boson fusion has the largest reach.   Trilepton searches and invisible Higgs searches are less powerful, but can give corroborating evidence.  However, such standard searches could become inefficient for these signals if the analyses adopt model dependent tools such as matrix elements and neural nets.  
}

\end{center}
\end{titlepage}


\section{Introduction}
\label{sec:intro}
Given its capabilities, the LHC will definitively explore the weak scale, especially any physics associated with electroweak symmetry breaking (EWSB). In weakly coupled models of EWSB, the breaking occurs through the vacuum expectation value of a scalar Higgs field.  All such models predict a new massive state, the Higgs boson, whose identification at the LHC is crucial. In the Standard Model, the couplings of the Higgs field are well known and its production and decay mechanisms well studied.   So far it has eluded searches with LEP II placing a bound on the Higgs mass near its kinematical limit and the Tevatron is still actively pursuing its own Higgs search.  However, ultimately this will be fully tested at the LHC,  whose reach for excluding or discovering a Standard Model Higgs extends over the entire relevant mass range. 

However, there are good reasons suggesting that the Higgs search will be far more difficult.  Such a scenario, often called a nonstandard Higgs (for a review, see \cite{nonstandardreview}), can be motivated in the following way.  First, the Higgs mass is suggested to be lighter than the LEP II bound of 114.4 GeV because both indirect constraints from electroweak precision tests and many theories beyond the Standard Model predict it to be near the $Z$ mass.  Second, for such masses, the Higgs Standard Model decay is mostly to $b$ quarks which, due to its small Yukawa coupling, can be dominated by Higgs decays into new light states.  If the new light states further decay, the direct search limits from LEP II can be significantly weaker.        
Such modifications diminish the prospects of discovering the Higgs in its Standard Model decays due to their reduced branching ratio.  Given the importance of the Higgs boson, it is therefore highly desirable to consider a broad range of possible decays for the Higgs in order to maintain discovery. 
LEP has explored many options for nonstandard Higgs decay, putting bounds that are in many cases as strong as Standard Model bounds.  In particular, two body decays to Standard Model particles are very well constrained, including invisible decays into neutrinos or any other neutral particle that escapes the detector.   However, nonstandard decays that involve three or more particles can be less constrained \cite{nonstandardreview}.  For instance, an intermediate scenario where the Higgs decays partly to visible particles and partly to invisible ones can allow a Higgs around $100$ GeV \cite{Chang:2007de}. 

Specifically, the following was considered, a Higgs decaying into two neutral particles $h \rightarrow \chi_2 \chi_1$, where $\chi_1$'s are collider stable and $\chi_2$ decays to $\chi_1$ plus two Standard Model fermions. A few interesting possibilities for the decay of $\chi_2$ were explored and their consequences for various LEP searches were examined \cite{Chang:2007de}. The first possibility is that $\chi_2$ decays through a three body decay to 2 leptons or 2 jets and a $\chi_1$. If the branching ratio to leptons is small enough, this mode could be allowed for Higgses of mass $\sim 100$ GeV.  Another interesting possibility is a two body decay of $\chi_2$ to $\chi_1$ and a scalar $\phi$. The scalar can subsequently decay to a pair of $\tau$'s or a pair of $b$'s, with the $b$ decays more highly constrained.   

In this paper we examine the possibility of discovering these types of nonstandard decay at the Tevatron and the LHC.  Most of the previous studies of nonstandard Higgs decays have motivated new analyses to be performed.  These looked for   
Higgs decays to 4$\tau$ \cite{Graham:2006tr, Forshaw:2007ra,:2008uu,Rottlander:2008zz,Belyaev:2008gj}, 4$b$ \cite{Cheung:2007un, Carena:2007jk}, 4$\mu$ \cite{Zhu:2006zv}, 4$\gamma$\cite{Chang:2006bw}, 2$\gamma$ + 2 jets \cite{Martin:2007dx},  multijets with displaced vertices \cite{Kaplan:2007ap}, four leptons with missing energy  \cite{Datta:2000ja}, and decays into right handed neutrinos which then decay with displaced vertices \cite{Graesser:2007yj,Graesser:2007pc} .      
In this note, we take a different approach.  Instead of a dedicated analysis, we restrict ourselves to existing Higgs and supersymmetry search topologies to estimate the prospect of seeing the nonstandard mode without having to design cuts that are specific to these decays.\footnote{A notable exception is the application of the 2$\gamma$ search in \cite{Dobrescu:2000jt}.}   We consider the following topologies:
\begin{itemize}
\item A  3 body decay of $\chi_2$ to $\chi_1$ and two charged leptons. This leads to the same  final states as a Standard Model Higgs decaying to 2 $W$'s that in turn decay leptonically.
\item Associated production of a Higgs with a $W$ which can lead to trileptons $+$ missing energy, which is the same topology as the famous supersymmetry signature of neutralino + chargino production. 
\item A $\chi_2$ decay to $\chi_1$ plus neutrinos, where the Higgs decay is invisible and is picked up by invisible Higgs searches.
\item A 2 body decay of $\chi_2$ to $\chi_1$ and $\phi$ which can lead to a signature similar to $h\rightarrow \tau^+ \tau^-$ which will also be looked for at the LHC.       
\end{itemize}
Of these modes, the first is known to offer good prospects for Standard Model Higgs discovery at the LHC for both light and intermediate Higgses produced in vector boson fusion.\footnote{This $h\to WW$ decay mode was searched for at the Tevatron, where it was recently used to rule out a Standard Model Higgs of $170$ GeV at $95\%$ C.L.  \cite{Bernardi:2008ee}.}  In fact, this search topology will have the highest reach for discovering such nonstandard decays.  The second and third modes are more limited in their discovery potential, but provide complementary information giving more credence to the nonstandard Higgs hypothesis.  Finally, the last mode is too difficult to detect even taking into account additional objects such as tagged jets or missing energy.  Most importantly, the first mode is a concrete example that nonstandard Higgs decays can be discovered not only in a standard analysis, but in early running at the LHC.  

The rest of the paper is organized as follows. In section \ref{3body}, we examine the 3 body decay of $\chi_2$ to leptons and missing energy. We estimate the efficiency of cuts originally designed to look  for a Standard Model Higgs decaying to $WW$ at Tevatron and LHC when applied to the nonstandard Higgs decay.  Given that this is the strongest discovery channel, we discuss possibilities of estimating the mass of the Higgs, and the difference of our signal from the background from supersymmetry signals. In a similar way, we calculate the efficiency of supersymmetry trilepton searches for a nonstandard Higgs produced in association with a $W$.  We also briefly discuss the prospect of seeing the invisible Higgs decay.  In section \ref{2body}, we consider the 2 body decay of $\chi_2$ and consider the possibility of discovering this nonstandard decay in $h \rightarrow \tau \bar{\tau}$ and $h \rightarrow b \bar{b}$ searches.  Finally, in section \ref{conclusions}, we conclude.

\section{Three body Decay of $\chi_2 \rightarrow l\bar{l} + \me$ \label{3body}}
\subsection{Tevatron Searches}
When $\chi_2$ decays through a 3 body decay to 2 leptons and $\chi_1$, the Higgs decay products are the same as that of a Standard Model Higgs decaying to two leptonic $W$'s (with the exception of the $e \mu$ mode which exist for $h\rightarrow WW$, but not for $h \rightarrow \chi_1 \chi_1 l^+ l^-$). At the Tevatron, $h \rightarrow W W$ is searched for in gluon fusion production, which has the largest cross section: $\sigma_{g g \rightarrow h}(m_h=160\ \text{GeV}) \times \text{Br}(h\rightarrow W W) = 240$ fb,  and is the most promising mode for the discovery of a heavy Higgs. The most recent analyses\cite{Aaltonen:2008ec,D0WWnew},  of this mode use neural net techniques which makes their reach for a different signal difficult to determine. To estimate the possible reach or exclusion of the nonstandard mode from the Tevatron, we use an older, cut based analysis of D\O \cite{D0WW} which looks in the $2 e$ + $\Sla{E_T}$ channel for a $m_h =120$ GeV Higgs by imposing the following cuts:
\begin{itemize}
\item 2 leptons with ${p_T}_1 > 15$ GeV, ${p_T}_2 >10$ GeV and $m_{l l} > 15$ GeV.
\item $\Sla{E_T} > 20 $ GeV
\item $m_{e e} < 60$ GeV
\item ${p_T}_1 + {p_T}_2 + \Sla{E_T} > 80$ GeV
\item ${m_T}_{\text{min}} > 45$ GeV, where $m_T = \sqrt{2 p_T^l \Sla{E_T} (1-\cos \Delta \phi)}$.
\item $H_T  = \sum_{\text{jets}} p_T <100$ GeV.
\item $\Delta \phi_{l l} < 2.0$
\end{itemize}
The largest potential background for this channel is Drell-Yan $Z/\gamma^* \rightarrow e^+ e^-$ which is suppressed by the $\Sla{E_T}$ cut. The remaining cuts shown above are optimized for $m_h=120$ GeV and also suppress background from $WW,WZ,ZZ,W+ \text{jets}$ and $t \bar{t}$. Once the cuts are applied, $WW, WZ$ and $W+ \text{jets}$ constitute the most important background.  Using this analysis on $\sim 950\ \text{pb}^{-1}$ of data, and combining a similar search in the $e-\mu$ channel, D\O \ sets a $95\%$ C.L. limit of
\begin{equation}
\label{d0bound}
\sigma_\text{SM}(g g \rightarrow h) \times \text{Br}(h \rightarrow W W) < 6.3 \ \text{pb}
\end{equation}
for $m_h = 120\ \text{GeV}$, while it is expected to be $0.08$ pb in the Standard Model. The efficiency of the cuts for a Standard Model Higgs of mass $120\ \text{GeV}$ is found to be $7 \%$ by \cite{D0WW}. Using Pythia \cite{Sjostrand:2006za} and PGS4\footnote{Available at http://www.physics.ucdavis.edu/$\sim$conway/research/software/pgs/pgs4-general.htm} for simulating the Standard Model Higgs signal  and detector effects, we find an efficiency of $\sim 12 \%$. For the nonstandard Higgs decay to  $l^+ l^- + \sla{E_T}$  the kinematics are different and  we find the efficiency of the cuts to be diminished. In particular, the $\Sla{E_T}$ and the cut on ${p_T}_1 + {p_T}_2 + \Sla{E_T}$ are found to be significantly less efficient. For example, choosing $m_h =100 \ \text{GeV}, m_{\chi_2} =50 \ \text{GeV}, m_{\chi_1}=10 \ \text{GeV}$, we find the overall efficiency to be  about $10$ times lower than for the $120$ GeV Standard Model Higgs.  Rescaling the bound of eq.(\ref{d0bound}) to take into account the difference in efficiency and the difference in branching ratio  leads to a bound on the nonstandard Higgs production of:
\begin{equation}
\sigma_\text{NS}(g g \rightarrow h) \times \text{Br}(h \rightarrow \chi_2 \chi_1) < 20 \ \text{pb} \times \frac{\text{Br}(Z \rightarrow e^+ e-)}{\text{Br}(\chi_2 \rightarrow e^+ e^- \chi_1)},
\end{equation}
where we expect $\sigma_\text{NS} \times \text{Br}(h \rightarrow \chi_2 \chi_1) \sim 1 \ \text{pb}$ if the production of nonstandard Higgs is the same as Standard Model one, and branching ratio of the Higgs to $\chi_2 \chi_1$ is close to $1$.  Other spectra can lead to larger efficiencies, for example choosing $m_h=110 \ \text{GeV}, m_{\chi_2} = 60 \ \text{GeV}$ and $ m_{\chi_1} = 10 \ \text{GeV}$ results in a factor of $3$ improvement, which is still however insufficient to put a significant bound on the nonstandard Higgs production. We therefore conclude that this mode is not excluded by the Tevatron, and that even with more statistics it would require a dedicated analysis, unless the branching ratio of $\chi_2$ to leptons is significantly larger than the $Z$ branching ratio to leptons.  A promising technique would be to flavor subtract, by subtracting final states with different flavor from final states with the same flavor. This should reduce backgrounds from $W$ production while having little effect on the nonstandard signal which only produces different flavor of leptons through leptonic $\tau$ decay.

\subsection{Higgs Production in Vector Boson Fusion at the LHC}

Decay of the Higgs to two $W$'s is also one of the preferred modes of discovery at the LHC for a Standard Model Higgs, even for somewhat light masses. It has been estimated by both the Atlas and CMS collaboration that by taking advantage of the vector boson fusion (VBF) production mechanism, this mode could  be seen above background with $30 \ \text{fb}^{-1}$ of data,  for Higgs masses between $140$ GeV and $190$ GeV. These searches  could in principle also see an excess from a nonstandard Higgs decaying to leptons and missing energy.  To assess the reach of planned LHC  searches for the nonstandard decay mode described above, we use the vector boson fusion Higgs study done for Atlas \cite{Asai:2004ws} \footnote{The newest Atlas study of this mode \cite{Aad:2009wy} uses more elaborate analysis techniques which make their analysis difficult to duplicate for our signal.}. The study of CMS \cite{Yazgan:2007cd} is slightly more pessimistic as they find a larger $t \bar{t}$ background.   Incidentally, as we will discuss later, the vector boson fusion production will help prevent this signal from being contaminated by other new physics signals like supersymmetry. 

Using Pythia, we generated signal events for the following process: $pp \rightarrow q q (h\rightarrow \chi_1 \chi_2 \rightarrow \chi_1 \chi_1 l^+ l^-)$, where $l$ stands for $e$, $\mu$ or $\tau$. We then used PGS4 as a detector simulation, and  applied the same cuts as in \cite{Asai:2004ws}:
\begin{itemize}
\item 2 leptons ($e$ or $\mu$) with $p_T(e) > 15 \ \text{GeV}$, $p_T(\mu) > 10 \ \text{GeV}$ and $\left | \eta \right| < 2.5$
\item 2 jets with large rapidity gap: ${p_T}_1 > 40 \ \text{GeV}$, ${p_T}_2 > 20 \ \text{GeV}, \left| \eta_1-\eta_2 \right| > 3.8$
\item $\Delta \phi_{l l} < 1.5$, $\Delta R_{l l} < 1.6$, $m_{l l} < 65 \ \text{GeV}$
\item $ 600 \ \text{GeV} < m_{j j}< 2500 \ \text{GeV}$
\item Transverse momentum balance: $\left| \mathbf {p_T}_{\text{total}} \right|= \left| \mathbf{p_T}_{\text{jets}} + \mathbf{p_T}_{\text{leptons}} + \mathbf{\sla{p_T}} \right| < 30 \ \text{GeV}$
\item Jet veto: no extra jet with $p_T > 20 \ \text{GeV}$ and $\left|\eta \right| < 3.2$
\item $m_T(l l \nu) = \sqrt{2 p_T(l l) \sla{E_T} (1-\cos\Delta \phi(l l ,\sla{\mathbf{p_T}}))} > 20$ GeV.
\end{itemize}
These cuts take advantage of the vector boson fusion production mechanism by requiring two high $p_T$, forward  jets with  a large rapidity gap. The jet veto cut is applied because VBF has relatively little extra radiation compared to QCD processes. This reduces background involving color exchange such as $t \bar{t}$ and $W W $ + jets.  The $\Delta \phi$ and $\Delta R$ cuts on leptons are applied because leptons from the Higgs signal are typically closer together than from $t \bar{t}$ and $WW$ backgrounds.  The total transverse momentum cut is correlated with the jet veto cut, and is useful to reject $t \bar{t}$ background. For the $ee$ and $\mu \mu$ channels, which are the only possible channels for the nonstandard Higgs, $Z/ \gamma^*$ + jets is an important source of background and can be controlled by cuts on $m_{l l}$ and on $m_T(l l \nu)$. The cut on the $p_T$ of the leptons corresponds to the $p_T$ threshold of the lepton pair trigger \cite{Asai:2004ws}. A significant fraction of events would also pass the single lepton trigger ($p_T > 25$ GeV for electron and $p_T > 20$ GeV for muons). 

\begin{figure}[h]
\begin{center}
\psfrag{labelx}{$\Delta \phi_{ll}$}
\includegraphics[width=7cm]{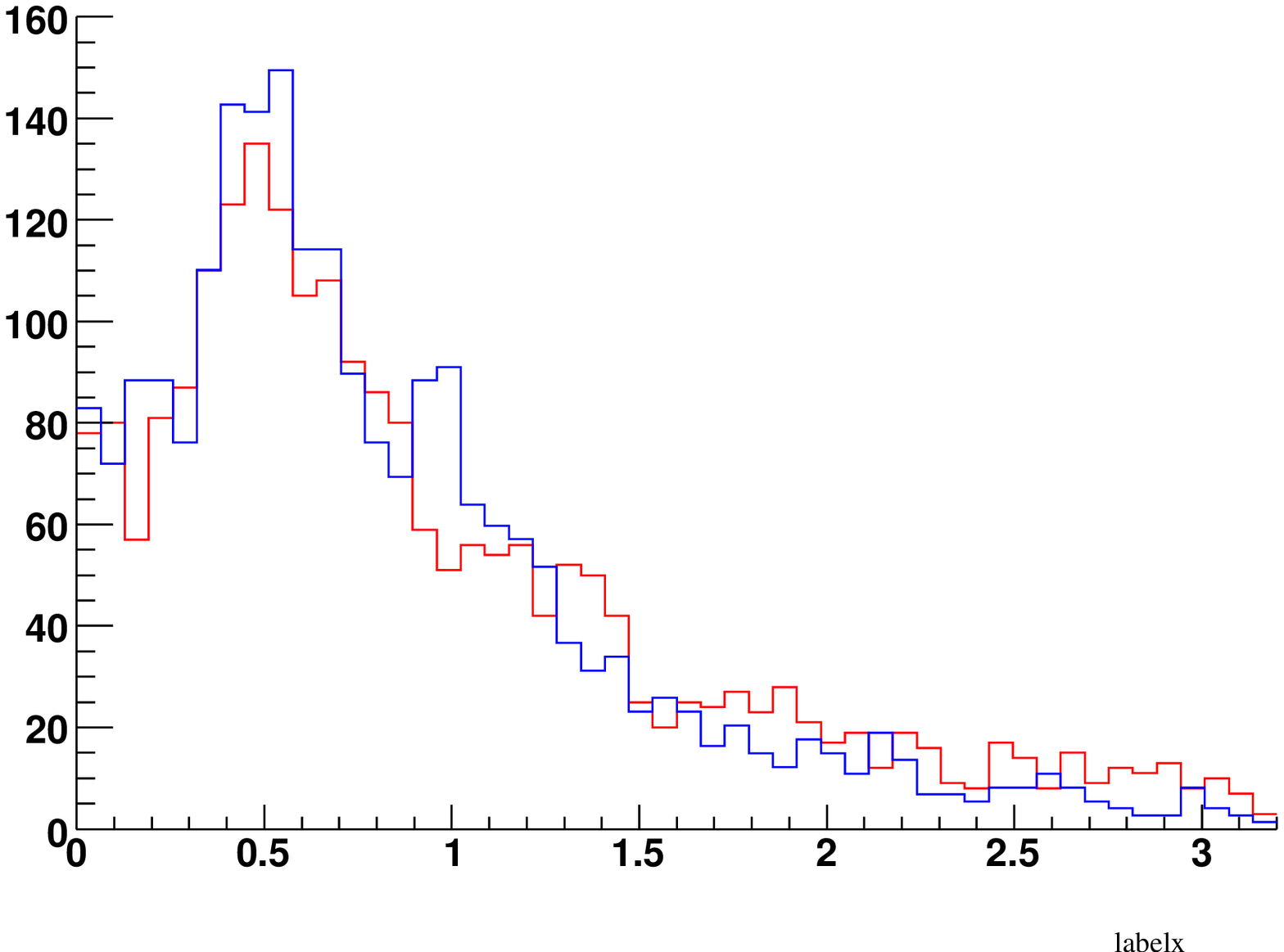}
\psfrag{labelx}{$p_T$(GeV)}
\includegraphics[width=7cm]{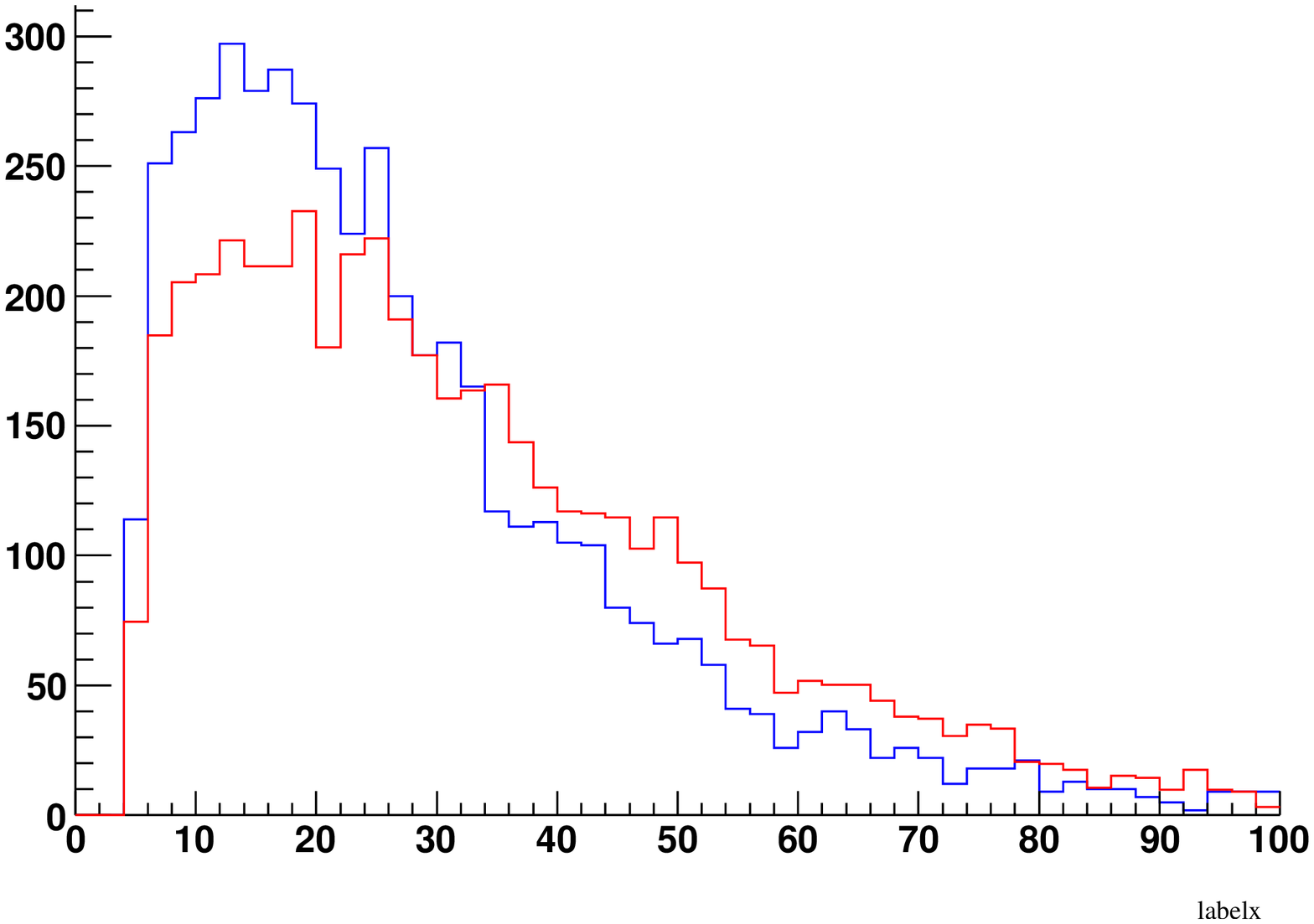}
\caption{Sample comparison of two kinematic variables for $h\to WW^*$ and the nonstandard Higgs decay.  Their similar shapes allow the original analysis to be efficient for the nonstandard signal.}
\label{kinematics}
\end{center}
\end{figure}

 When applied on events from the nonstandard decay $h \rightarrow \chi_2 \chi_1 \rightarrow \chi_1 \chi_1 l^+ l^-$ (where $l=e,\mu,\tau$), we find that the kinematics are similar enough that we retain a relatively large fraction of events (see figure \ref{kinematics}).   Choosing  $m_h = 100 \ \text{GeV}, m_{\chi_2} = 50 \ \text{GeV and } m_{\chi_1} = 10 \ \text{GeV}$, we find an efficiency of $\sim 1\%$ (note that this number does not include the $\chi_2$ branching ratio to leptons).  Assuming production of the nonstandard Higgs to be the same as the Standard Model Higgs, this would yield a  signal cross section of 
\begin{equation}
\label{atlassignal}
5 \ \text{fb} \times \frac{\text{Br}(\chi_2 \rightarrow l^+ l^- \chi_1)}{.10}\times  \text{Br}(h\rightarrow \chi_2 \chi_1)
\end{equation}
where we have normalized the leptonic branching ratio of $\chi_2$ to the $Z$'s, and did not take into account any triggering efficiency.
The backgrounds for this process  consist of $t \bar{t}$, $W W +$ jets and $\gamma^*/Z +$ jets events, and was found to be $1.33$ fb\cite{Asai:2004ws} for the $ e^+  e^- + \mu^+ \mu^-$ channel. This leads us to conclude that the nonstandard Higgs decay could easily be seen above background, even with a search originally designed to discover a Standard Model Higgs.
 Note that the background and signal cross section were estimated using the leading order calculation, and at next to leading order (NLO), QCD background could be enhanced relative to the signal which does not get a very large correction.  Ultimately, the background will be determined from data, but given our assumptions, that the nonstandard Higgs is produced with Standard Model cross section, it's branching ratio to $\chi_2 \chi_1$ is of order one, and the branching ratio of $\chi_2$ to leptons is the same as the $Z$, the large value of the expected signal (eq. \ref{atlassignal}) makes our conclusion rather insensitive to variations in the background estimate. 
  
For comparison, a Standard Model Higgs of $120$ GeV, would give a  signal that was estimated by \cite{Asai:2004ws} to be about $0.5$ fb. The cuts on the nonstandard  signal are $\sim 60 \%$ less efficient  than when applied to the $120$ GeV Standard Model Higgs; however, the light Standard Model Higgs has a small branching ratio to $WW$, while the nonstandard Higgs can have an overall larger rate into leptons. As a cross check of our method, we also generated the $120$ GeV Standard Model Higgs signal with Pythia and PGS4, and compared with \cite{Asai:2004ws}. We found our results to be within $20\%$, and consider this discrepancy as reasonable given that the detector simulation that we used is different than the one used in \cite{Asai:2004ws}, which used ATLFAST. 

\begin{figure}[h]
\begin{center}
\psfrag{labelx}{$m_{\chi_2} - m_{\chi_1}$(GeV)}
\includegraphics[width=11cm]{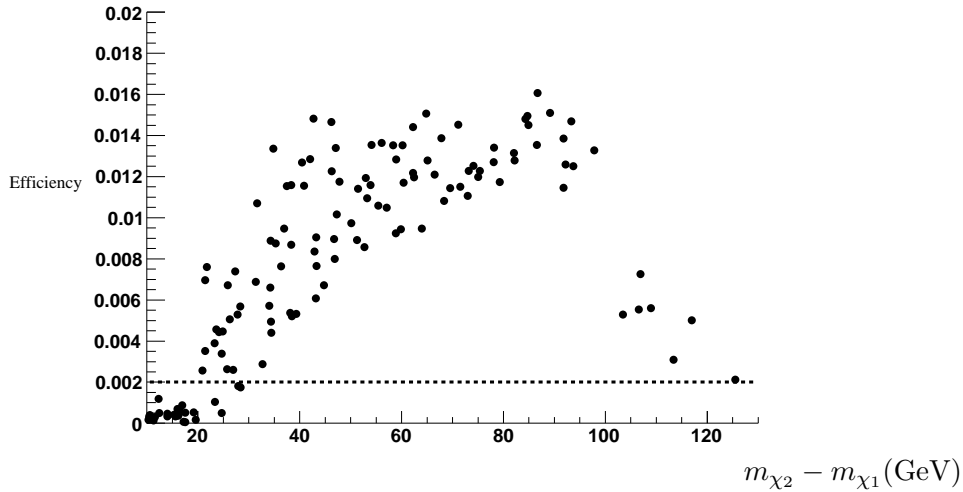}
\caption{The efficiency for the analysis cuts of $h\to WW^*$ on the nonstandard Higgs signal as a function of the mass difference between $\chi_2$ and $\chi_1$.  For reference, the dotted line is the efficiency required for a 5$\sigma$ discovery assuming 30 fb$^{-1}$, $\text{Br}(h \rightarrow \chi_2 \chi_1)=1$, and $\text{Br}(\chi_2 \rightarrow \chi_1 l^+ l^-) = \text{Br}(Z \rightarrow l^+ l^-) = .10$.}
\label{fig: effvsmassdiff}
\end{center}
\end{figure}

   However, the efficiency for the signal depends strongly on the mass difference between $\chi_1$ and $\chi_2$. As $\chi_2$ and $\chi_1$ become degenerate in mass, the leptons are soft and do not pass basic acceptance cuts. If on the other hand, the mass difference is close to the $Z$ mass, events are rejected because of the cut on $m_{ll}$.  We generated different spectrum for our signal, keeping the Higgs light ($m_h < 130 \ \text{GeV}$), and found that the efficiency of the cuts depends mostly on the $\chi_2 - \chi_1$ mass difference (see figure \ref{fig: effvsmassdiff}).  We also show on this plot the efficiency that would be needed to have $S/\sqrt{B} > 5$ with $30 \ \text{fb}^{-1}$ data, assuming the production cross section of a $100$ GeV Standard Model Higgs, $\text{Br}(h \rightarrow \chi_2 \chi_1)=1$, and $\text{Br}(\chi_2 \rightarrow \chi_1 l^+ l^-) = \text{Br}(Z \rightarrow l^+ l^-) = .10$. 

As the mass difference between $\chi_2$ and  $\chi_1$ decreases, the Higgs will decay invisibly, and this can also be seen in the vector boson fusion channel. The invisible channel can also be useful with larger mass difference as $\chi_2$ could have significant branching ratio to neutrinos. Studies of invisible Higgs decay \cite{Eboli:2000ze} have showed that at the LHC, it is sensitive at the 2$\sigma$ to a Higgs with an invisible  branching ratio of about $15 \%$ with 10 fb$^{-1}$.   However, the recent Atlas study \cite{Aad:2009wy} shows that taking into account systematic uncertainties, with 30 fb$^{-1}$, it would require a 60\% ``invisible'' branching ratio for a 2$\sigma$ effect.  These systematics will have to be overcome to be able to see evidence for this branching ratio.  

\subsubsection*{Determination of the Spectrum}
\begin{figure}[h]
\begin{center}
\psfrag{label}{$m_{l l}$(GeV)}
\includegraphics[width=8cm]{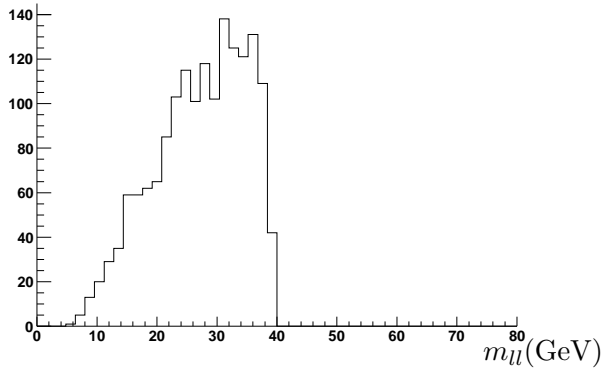}
\caption{A sample $m_{ll}$ distribution which exhibits a sharp edge near the mass difference $m_{\chi_2}-m_{\chi_1} = 40$ GeV.}
\label{fig: mll}
\end{center}
\end{figure}

\begin{figure}[h]
\begin{center}
\psfrag{labelx}{$m_{l l}$(GeV)}
\psfrag{labely}{$m_{\nu \nu}$ }
\includegraphics[width=8cm]{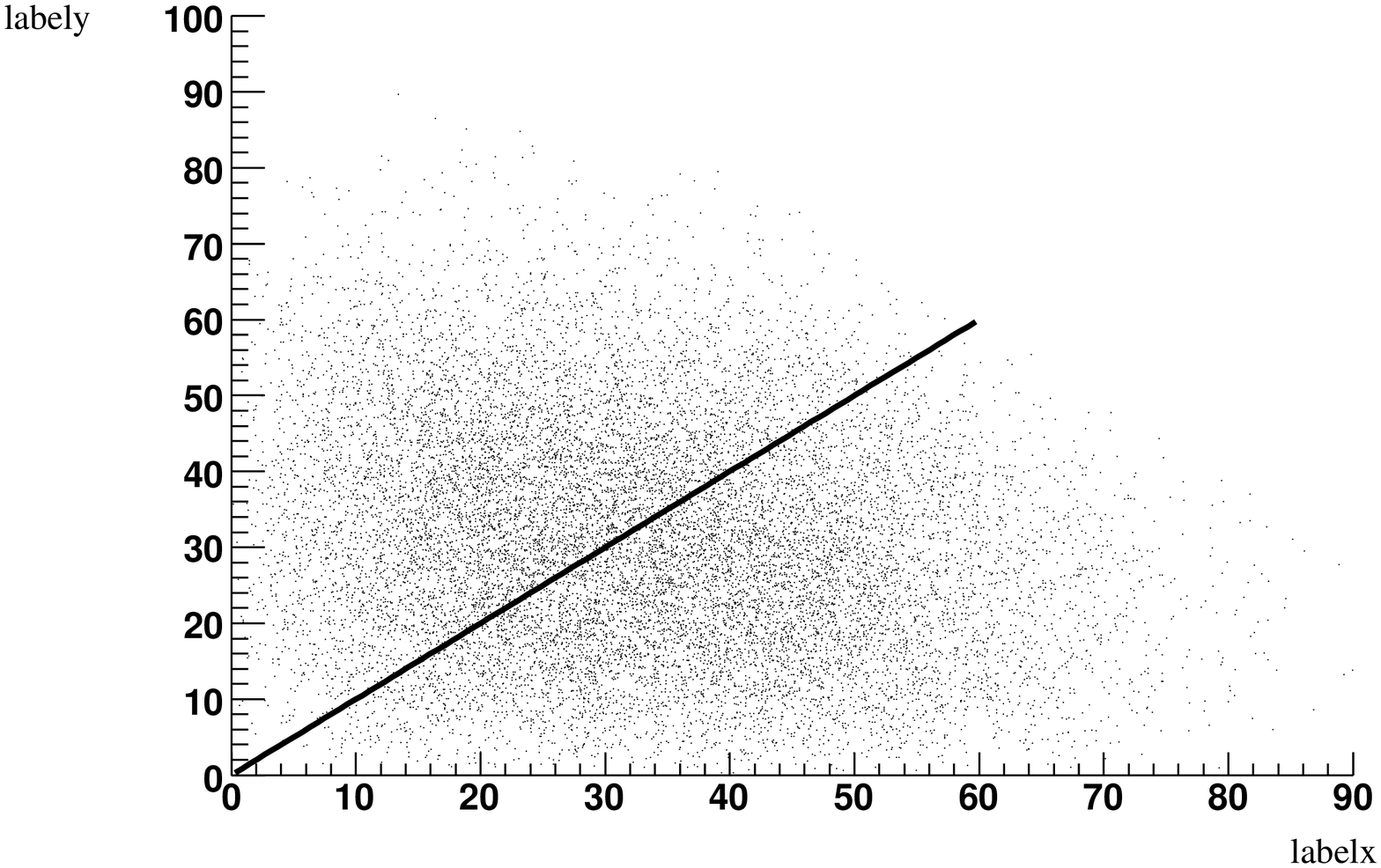}
\psfrag{labelx}{$m_{l l}$(GeV)}
\psfrag{labely}{$m_{\chi \chi}$ }
\includegraphics[width=8cm]{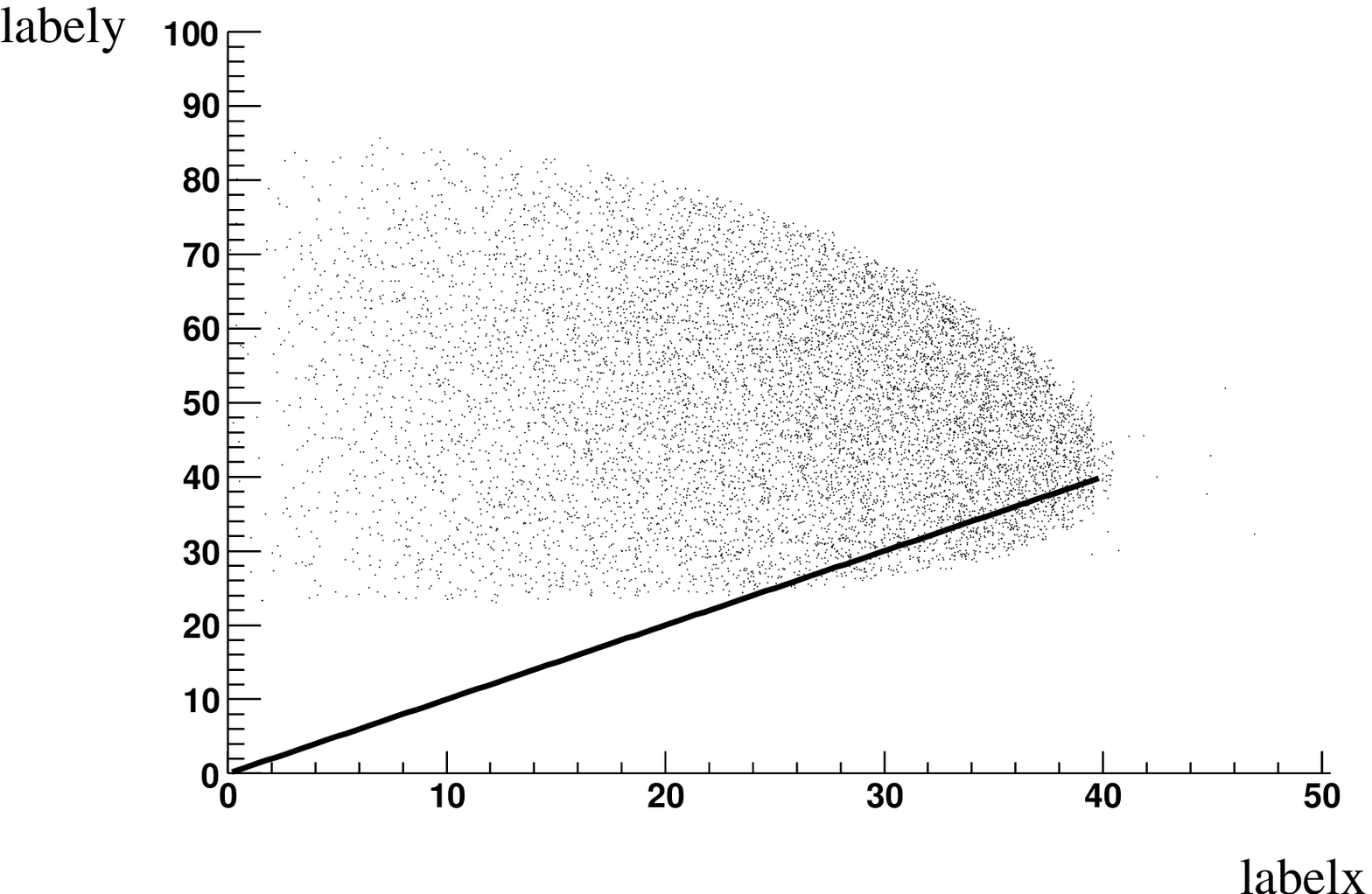}
\caption{$m_{ll}$ vs $m_{\text{invisible}}$ for a Standard Model Higgs of $120$ GeV (left) and a nonstandard Higgs of $100$ GeV (right).}
\label{fig: mllvsmchichi}
\end{center}
\end{figure}

\begin{figure}[h]
\begin{center}
\psfrag{labelx}{$m_T$(GeV)}
\includegraphics[width=8cm]{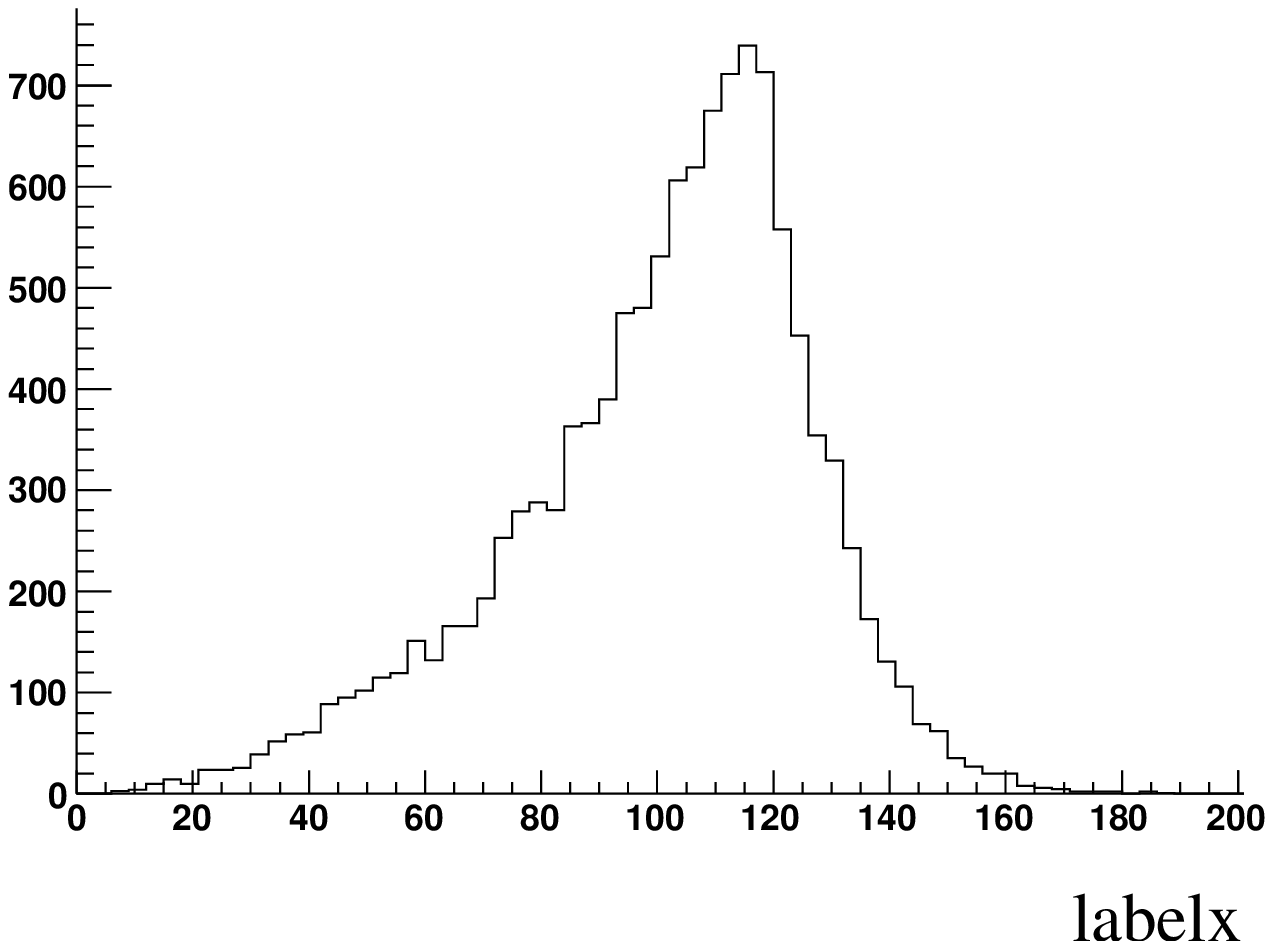}
\psfrag{labelx}{$m_T$(GeV)}
\psfrag{label40}{$m_{\chi_1} = 40$ GeV}
\psfrag{label20}{$m_{\chi_1} = 20$ GeV}
\psfrag{label0}{$m_{\chi_1} = 0$ GeV}
\includegraphics[width=8cm]{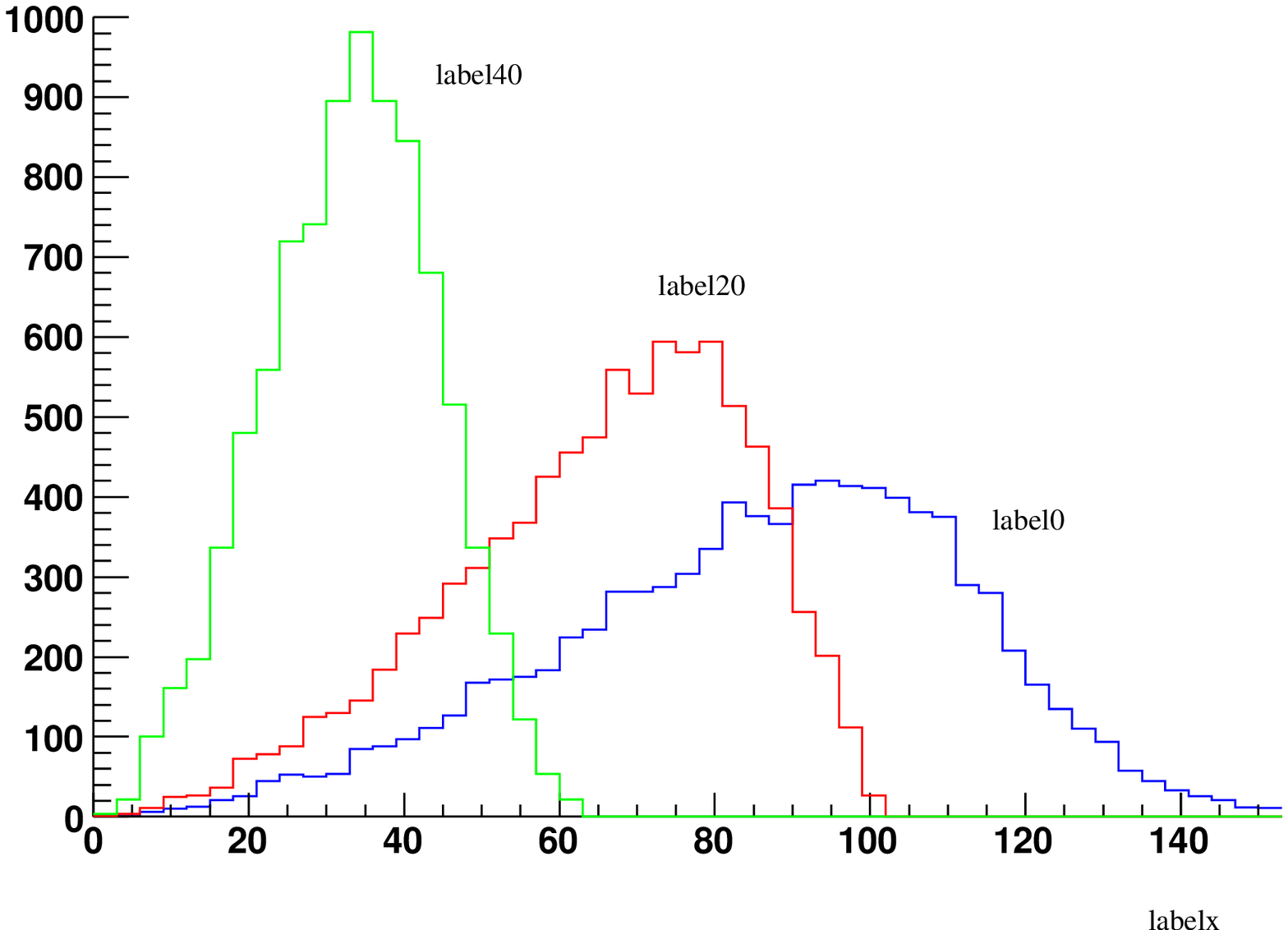}
\caption{Transverse mass $m_T$ defined in equation \ref{eq: mtr} for a Standard Model Higgs of $120$ GeV (left) and a nonstandard Higgs decay, with $m_h =110$ GeV, $m_{\chi_2}= 60$ GeV and different $\chi_1$ masses.}
\label{fig: tramss}
\end{center}
\end{figure}

If an excess of events is seen in the lepton channel, one would like to reconstruct the decay chain and ideally get an estimate of the Higgs mass.  The mass difference between $\chi_2$ and $\chi_1$ could  be easily read off the endpoint of the $m_{l l}$ distribution (see fig. \ref{fig: mll}). Determining the mass of the Higgs is however not as easy.
In the case of a Standard Model Higgs decaying  to $W W$, the Higgs mass  can be estimated by plotting the transverse mass of the leptons and missing energy system:
\begin{equation}
{m^2_T}_{W W} =\left( \sqrt{m_{l l}^2 + {p_{ll}}_T^2 } + \sqrt{m_{ll}^2+\sla{p_T}^2} \right)^2 - ({\mathbf p_{ll}}_T + \sla{\mathbf p_T})^2
\label{eq: mtr}
\end{equation}
While this distribution exhibits a jacobian peak at the Higgs mass \cite{Rainwater:1999sd,Kauer:2000hi},  it relies on the approximation $m_{ll} = m_{\nu \nu}$(see \cite{Barr:2009mx} for a new method that do not use this approximation). This is strictly true only when the two $W$'s are produced at rest, but as shown in figure \ref{fig: mllvsmchichi} it holds on average.  For the nonstandard  signal however, it is not the case (see figure \ref{fig: mllvsmchichi}), and the $m_T$ distribution doesn't have a peak at $m_h$ as shown in figure \ref{fig: tramss}.  If we assume that $\chi_2$ and $\chi_1$ are produced at rest, we obtain $m^2_{\chi \chi} =   2\left(m_{\chi_1}^2 + m_{\chi_1} \sqrt{m_{\chi_1}^2 + p_{ll}^2}\right)$. The  mass distribution based on this relation is more successful in estimating the Higgs mass,  and the situation becomes more similar to the Standard Model one. However it requires a knowledge of $m_{\chi_1}$ and with that information, it might be possible to get a better determination of the Higgs mass by constructing templates of the $m_T$ distribution for different Higgs masses, which can then be determined by fitting to the observed distribution.  

\subsubsection*{Background from Supersymmetry}
Given the nonstandard nature of the decay we are studying, if it were to cause an excess of events seen in the $h \rightarrow W W$ search,  it might not be  straightforwardly attributed to a Higgs decay. The absence of $e \mu$ events would indicate that the anomalous signal is not coming from $h \rightarrow W W$; however,  the presence of missing energy might point towards a supersymmetric decay. 
Neutralino or slepton pairs produced in vector boson fusion would give exact same final state as our decay. However production cross sections are in the attobarn range and are too small to be relevant \cite{Cho:2006sx}. 

\begin{figure}[h]
\begin{center}
\psfrag{labelx}{$\Sla{E}_T$(GeV)}
\psfrag{label100}{$m_h = 100$ GeV}
\psfrag{label200}{$m_{\tilde{q}} = 200$ GeV}
\psfrag{label500}{$m_{\tilde{q}}=500 $ GeV}
\includegraphics[width=8cm]{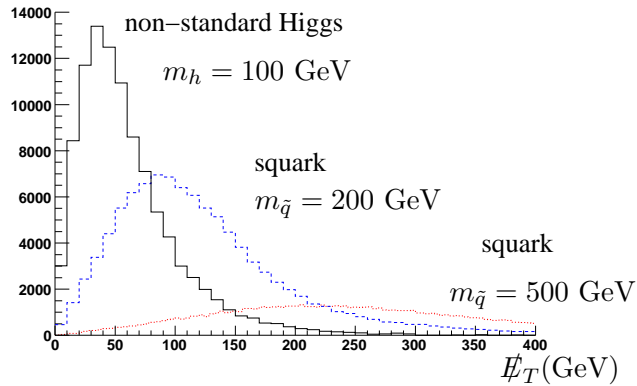}
\caption{The $\Sla{E}_T$ distribution for a nonstandard Higgs and squarks with mass 200 and 500 GeV.}
\label{met}
\end{center}
\end{figure}

Squark pair production can also produce a final state with 2 leptons $+$ 2 jets $+ \Sla{E_T}$.  For example, production of $\tilde{q}_L \tilde{q}_R$ followed by  $\tilde{q}_L \rightarrow \chi_2 q$, $\tilde{q}_R \rightarrow \chi_1 q$, can produce the two neutralinos and two jets. The jet kinematics would be quite different from the vector boson fusion process, however some events are expected to pass the cuts. We generated  $500$ GeV squarks decaying as above with $m_{\chi_2} = 50$ GeV and $m_{\chi_1} = 10$ GeV, and $\chi_2$ decaying to 2 leptons and $\chi_1$. We found that the Atlas cuts described in the previous section have an efficiency of $0.03 \%$ while the cross section for this process is $\sim 2 \ \text{pb}$ at leading order.  By interpreting a Higgs decay, $h \rightarrow \chi_2 \chi_1$ as a squark production process, we would conclude on a production cross section for squark that is $\sim 70$ times larger than the Higgs production cross section . The same process with squarks of $200$ GeV, would yield an efficiency of  $0.01\%$ with a production cross section of $\sim 20$ pb. However, given that we don't know the branching ratio of $\chi_2$ to leptons, such  simple counting experiments might not be enough to rule out squark production. For example, a nonstandard Higgs signal with low branching fraction to leptons might yield the same number of events as squark production with a large branching ratio to leptons. One could in addition look at various distributions to distinguish the two cases. For example, $\Sla{E}_T$ is  larger for squark production than for the Higgs process.  In figure \ref{met} we show the missing energy distribution for a nonstandard Higgs of $100$ GeV, and for  squarks of $500$ and $200$ GeV. 

If the nonstandard Higgs decay arises because it is truly part of a supersymmetric theory, where $\chi_2$ and $\chi_1$ are actual neutralinos, processes with squark and gluino will occur in addition to the Higgs signal. Seeing these additional signals might help to reconstruct the Higgs decay, and vice-versa.  For instance, the supersymmetric signals may shed light on the $\chi_2$ branching ratio and potentially the mass of $\chi_1$.   These issues make exploration of a specific supersymmetric scenario with its cross correlation between supersymmetry and Higgs signals extremely interesting, but is beyond the scope of this paper. 

\subsection{Trileptons}

When produced in association with a $W$, the nonstandard decay $h \rightarrow \chi_2 \chi_1 \rightarrow \chi_1 \chi_1 l^+ l^-$ can lead to events with 3 leptons and missing energy.  Trilepton events with missing energy also appear in supersymmetric theories from production of a neutralino and a chargino, and a variety of searches have been designed to look for them. Here we would like to estimate the possibility of seeing a nonstandard Higgs signal in these type of searches. In \cite{Sullivan:2008ki}, the authors revisit  proposed Atlas and CMS searches for trileptons, concentrating on background from heavy flavor. They found that semi-leptonic decay of a heavy flavor jet can yield isolated leptons and contribute significantly to the background. To reduce it, they  add a set of angular cuts on the leptons. We follow their analysis which asks for three leptons, missing energy, and imposes a jet veto before imposing the angular cuts on the leptons.  In more detail, the cuts are:
\begin{itemize}
\item Same flavor, opposite sign lepton pair with $p_T > 15 \ \text{GeV}(e)$, $p_T > 10 \ \text{GeV}$ ($\mu$), and $\left| \eta \right| < 2.5$.
\item $\left| m_{l l}-m_Z \right| > 10 \ \text{GeV}$.
\item $\Delta R_{l l} < 2.6$.
\item third lepton, with  $p_T > 15 \ \text{GeV}(e)$, $p_T > 10 \ \text{GeV}$ ($\mu$), and $\left| \eta \right| < 2.5$.
\item $\Sla{E} > 30 \ \text{GeV}$.
\item no extra jet with $p_T > 20 \ \text{GeV}$ and $\left| \eta \right| < 2.4$.
\item $\theta_{12} > 45^{\circ}, \theta_{13} > 40^{\circ}, \theta_{23} < 160^{\circ}$.
\end{itemize}
The first 6 cuts are copied from the Atlas search, and the last one represents a set of angular cuts applied to the $p_T$ ordered leptons in the center of mass frame of the 3 leptons. We found  the efficiency of these cuts for our signal  to be $5\%$, while the background was estimated at $35$ fb in \cite{Sullivan:2008ki}.  Thus, $S/\sqrt{B} > 2, 3, 5$ would require luminosities of $60, 130, 370$ fb$^{-1}$, where we have assumed the standard $hW$ cross section of 2.2 pb and $\chi_2$ branching ratios to electrons and muons to be set by the $Z$'s.

Even though this mode is hard to discover, it can provide complementary information by giving an additional 2-4$\sigma$ evidence for the scenario in the high luminosity run. 
\section{ Two body Decay of $\chi_2 \rightarrow \phi \chi_1$ \label{2body}}
\subsection{$\tau \tau$ + missing energy}
Another interesting possibility for the decay of $\chi_2$ is a two body decay  to $\chi_1$ and a scalar $\phi$. Depending on its mass, we can imagine $\phi$ decaying to $b \bar{b}$ or, if it's lighter, to $\tau^+ \tau^-$.  In the case of the $\tau^+ \tau^-$ decay, the final state would be the same as $h \rightarrow \tau^+ \tau^-$, which is a promising discovering mode for a light Higgs produced in vector boson fusion.  This can be searched for, either with both $\tau$'s decaying leptonically, or with one $\tau$ going leptonically and the other to a hadronic mode. The strategy is similar to the $WW$ search. For the purely leptonic mode, one looks for 2 hard jets with large separation in $\eta$, and for two leptons. Various kinematic cuts are then applied to the leptons and jets, and a jet veto is imposed to reduce QCD backgrounds. Next, the $\tau$'s momenta are reconstructed by assuming that the neutrinos are collinear with the leptons, and a cut on $m_{\tau \tau}$ is applied \cite{Asai:2004ws}.   

In principle, before any cuts, the nonstandard Higgs decay could produce a larger rate than the Standard Model Higgs, if $\phi$ has a large branching ratio to $\tau \tau$. This could happen for example if $\phi$ couples to fermions proportionally to their mass like the Higgs, but was lighter than  $2 m_b$. However, if $\phi$ is too light its decay product will tend to be collimated and/or soft.%
For a scalar $\phi$ of $8$ GeV, just below the $b\bar{b}$ threshold and a $100$ GeV Higgs, we find  the efficiency of the cuts to be smaller than $10^{-5}$, which would lead to a signal of less than $0.006$ fb.   When the mass of $\phi$ is increased the acceptance for leptons is also increased. Ultimately, the main hurdle for the discovery of this mode becomes the reconstruction of the $\tau$'s. Because of additional missing energy from $\chi_1$, the $\tau$ four-momenta and consequently $m_{\tau \tau}$ cannot be reconstructed. Therefore the cut on $m_{\tau \tau}$, which is imposed to reduce the  $Z +$ jets background is less efficient, and the $\phi$ mass peak cannot be reconstructed. We find that taking the branching ratio of $\phi$ to $\tau$'s to be the same as a Standard Model Higgs, this mode cannot be detected even with $300 \ \text{fb}^{-1}$ of data.  Thus,  this mode will not be seen in this planned search. 

\subsection{$\phi \rightarrow b \bar{b}$}

If $\phi$ decays to a $b$ pair, the signal would look similar to a Standard Model Higgs decaying to $b \bar{b}$. This mode is notoriously difficult at the LHC. We therefore examine in more detail the prospects at the Tevatron. The final state of the nonstandard decay is 2 $b$'s and missing energy, which is the same as a Standard Model Higgs produced in association with a $Z$, with the Higgs decaying to $b \bar{b}$ and the $Z$ to neutrinos. The possible advantage of the nonstandard Higgs is that we can get this final state in gluon fusion production which has a larger cross section  than associated production. Unfortunately the missing energy for the nonstandard Higgs produced in gluon fusion is rather small, and the efficiency of the cuts used by CDF to look for $\sla{E}_T + b \bar{b}$ are a factor of $100$ less efficient on our signal, while the gain in production and branching ratio is only $15$ \cite{Aaltonen:2008mi}.   Therefore, it is unlikely that this Tevatron analysis will be sensitive to this nonstandard decay mode. 

\section{Conclusions \label{conclusions}}
In this note, we have explored the possibility of detecting a Higgs that decays in a nonstandard topology, partly to missing energy and partly to visible particles. In particular, we consider the decay $h \rightarrow \chi_2 \chi_1$ where $\chi_1$ is stable and $\chi_2$ decays back to $\chi_1$ and visible particles. This could arise for example in the NMSSM, the MSSM with right handed sneutrinos, or in models with a heavy neutrino. Among various possible decays of $\chi_2$, its decay to leptons and missing energy seems the most promising. While we find that the Tevatron has little chance of seeing this mode in its existing searches for a Standard Model Higgs, we find that LHC searches could be sensitive to this type of decay. 

Our approach has been to follow existing Higgs and supersymmetry searches and apply the same cuts on a nonstandard Higgs signal, that we generated using PYTHIA and PGS, that has the same topology.   The most recent Higgs studies done for Atlas, and presented by the Tevatron experiments use sophisticated analyses techniques such as Neural Nets and matrix elements. While very powerful for the specific model they are set to test, they are not well suited for our purposes as estimating their efficiency for a different signal is difficult.  In the worst case scenario, these signal dependent choices can even make these searches insensitive to nonstandard decays with the same event topology.  

Yet, as our study shows, cut based searches offer the possibility of discovering more than what they were originally designed for, and as such are more model independent.   Of these standard analyses, the search for vector boson fusion produced $h\to WW^*$ was shown to be capable of detecting the nonstandard Higgs with only 30 fb$^{-1}$ for most of the parameter space.  We explored some methods of mass reconstruction, but determined that a precise method requires knowing the mass of the particle $\chi_1$, which could be measured from other new physics signals.   Two other standard analyses, the search for invisible Higgs decays and the supersymmetric trilepton search, were shown to be weaker but able to give corroborating 2-3$\sigma$ evidence for these decays with 100 fb$^{-1}$.   Finally, the case of two body decays $\chi_2 \to \phi \chi_1$ where the scalar $\phi$ decays into $b$ quarks or $\tau$ leptons was shown to be too difficult to be detected at the LHC or Tevatron.  Better ideas will be needed to search for this possibility.   

More generally, this study shows that LHC's discovery potential is more powerful than just focusing on particular signals.  If the analyses are generic enough so as not to exclude other signals with similar event topology, these analyses can discover more than their original motivation.  This gives us additional hope that the LHC will discover the new physics of the TeV scale, whatever it turns out to be.  

\section*{Acknowledgements}
The work of SC was supported in part by NSF CAREER grant PHY-0449818 and DOE grant \# DE-FG02-06ER41417.  TG is supported by a SUPA advanced fellowship. SC  acknowledges the hospitality and support of the Kavli Institute for Theoretical Physics China, CAS, Beijing 100190, China, where some of this work was undertaken. 

\bibliography{higgsMET}

\providecommand{\href}[2]{#2}\begingroup\raggedright\begin{thebibliography}{10}

\bibitem{nonstandardreview}
S.~Chang, R.~Dermisek, J.~F. Gunion, and N.~Weiner, ``{Nonstandard Higgs Boson
  Decays},''
  \href{http://dx.doi.org/10.1146/annurev.nucl.58.110707.171200}{{\em Ann. Rev.
  Nucl. Part. Sci.} {\bf 58} (2008)  75--98},
\href{http://arxiv.org/abs/0801.4554}{{\tt arXiv:0801.4554 [hep-ph]}}.

\bibitem{Chang:2007de}
S.~Chang and N.~Weiner, ``{Nonstandard Higgs Decays with Visible and Missing
  Energy},'' \href{http://dx.doi.org/10.1088/1126-6708/2008/05/074}{{\em JHEP}
  {\bf 05} (2008)  074},
\href{http://arxiv.org/abs/0710.4591}{{\tt arXiv:0710.4591 [hep-ph]}}.

\bibitem{Graham:2006tr}
P.~W. Graham, A.~Pierce, and J.~G. Wacker, ``Four taus at the tevatron,''
\href{http://arxiv.org/abs/hep-ph/0605162}{{\tt hep-ph/0605162}}.

\bibitem{Forshaw:2007ra}
J.~R. Forshaw, J.~F. Gunion, L.~Hodgkinson, A.~Papaefstathiou, and A.~D.
  Pilkington, ``{Reinstating the 'no-lose' theorem for NMSSM Higgs discovery at
  the LHC},'' \href{http://dx.doi.org/10.1088/1126-6708/2008/04/090}{{\em JHEP}
  {\bf 04} (2008)  090},
\href{http://arxiv.org/abs/0712.3510}{{\tt arXiv:0712.3510 [hep-ph]}}.

\bibitem{:2008uu}
N.~E. Adam {\em et al.}, ``{Higgs Working Group Summary Report},''
\href{http://arxiv.org/abs/0803.1154}{{\tt arXiv:0803.1154 [hep-ph]}}.

\bibitem{Rottlander:2008zz}
I.~Rottlander, ``{Development of a benchmark parameter scan for Higgs bosons in
  the NMSSM model and a study of the sensitivity for H -- > AA --> 4tau in
  vector boson fusion with the ATLAS detector},''. CERN-THESIS-2008-064.

\bibitem{Belyaev:2008gj}
A.~Belyaev {\em et al.}, ``{The Scope of the 4 tau Channel in Higgs-strahlung
  and Vector Boson Fusion for the NMSSM No-Lose Theorem at the LHC},''
\href{http://arxiv.org/abs/0805.3505}{{\tt arXiv:0805.3505 [hep-ph]}}.

\bibitem{Cheung:2007un}
K.-m. Cheung, J.~Song, and Q.-S. Yan, ``Roles of higgs decay into two
  pseudoscalar bosons in the search of intermediate-mass higgs boson,''
\href{http://arxiv.org/abs/arXiv:0710.1997 [hep-ph]}{{\tt arXiv:0710.1997
  [hep-ph]}}.

\bibitem{Carena:2007jk}
M.~Carena, T.~Han, G.-Y. Huang, and C.~E.~M. Wagner, ``Higgs signal for h to aa
  at hadron colliders,''
\href{http://arxiv.org/abs/arXiv:0712.2466 [hep-ph]}{{\tt arXiv:0712.2466
  [hep-ph]}}.

\bibitem{Zhu:2006zv}
S.-h. Zhu, ``{Unique Higgs boson signature at colliders},''
\href{http://arxiv.org/abs/hep-ph/0611270}{{\tt arXiv:hep-ph/0611270}}.

\bibitem{Chang:2006bw}
S.~Chang, P.~J. Fox, and N.~Weiner, ``Visible cascade higgs decays to four
  photons at hadron colliders,'' {\em Phys. Rev. Lett.} {\bf 98} (2007)
  111802,
\href{http://arxiv.org/abs/hep-ph/0608310}{{\tt hep-ph/0608310}}.

\bibitem{Martin:2007dx}
A.~Martin, ``Higgs cascade decays to gamma gamma + jet jet at the lhc,''
\href{http://arxiv.org/abs/hep-ph/0703247}{{\tt hep-ph/0703247}}.

\bibitem{Kaplan:2007ap}
D.~E. Kaplan and K.~Rehermann, ``Proposal for higgs and superpartner searches
  at the lhcb experiment,'' {\em JHEP} {\bf 10} (2007)  056,
\href{http://arxiv.org/abs/arXiv:0705.3426 [hep-ph]}{{\tt arXiv:0705.3426
  [hep-ph]}}.

\bibitem{Datta:2000ja}
A.~Datta, P.~Konar, and B.~Mukhopadhyaya, ``{New Higgs signals from vector
  boson fusion in R-parity violating supersymmetry},''
  \href{http://dx.doi.org/10.1103/PhysRevD.63.095009}{{\em Phys. Rev.} {\bf
  D63} (2001)  095009},
\href{http://arxiv.org/abs/hep-ph/0009112}{{\tt arXiv:hep-ph/0009112}}.

\bibitem{Graesser:2007yj}
M.~L. Graesser, ``Broadening the higgs boson with right-handed neutrinos and a
  higher dimension operator at the electroweak scale,'' {\em Phys. Rev.} {\bf
  D76} (2007)  075006,
\href{http://arxiv.org/abs/arXiv:0704.0438 [hep-ph]}{{\tt arXiv:0704.0438
  [hep-ph]}}.

\bibitem{Graesser:2007pc}
M.~L. Graesser, ``Experimental constraints on higgs boson decays to tev-scale
  right-handed neutrinos,''
\href{http://arxiv.org/abs/arXiv:0705.2190 [hep-ph]}{{\tt arXiv:0705.2190
  [hep-ph]}}.

\bibitem{Dobrescu:2000jt}
B.~A. Dobrescu, G.~L. Landsberg, and K.~T. Matchev, ``Higgs boson decays to
  cp-odd scalars at the tevatron and beyond,'' {\em Phys. Rev.} {\bf D63}
  (2001)  075003,
\href{http://arxiv.org/abs/hep-ph/0005308}{{\tt hep-ph/0005308}}.

\bibitem{Bernardi:2008ee}
{\bf Tevatron New Phenomena Higgs Working Group} Collaboration, G.~Bernardi
  {\em et al.}, ``{Combined CDF and D0 Upper Limits on Standard Model Higgs
  Boson Production at High Mass $(155-200-GeV/c^{2)}$ with 3 $fb^{-1}$ of
  data},''
\href{http://arxiv.org/abs/0808.0534}{{\tt arXiv:0808.0534 [hep-ex]}}.

\bibitem{Aaltonen:2008ec}
{\bf CDF} Collaboration, T.~Aaltonen {\em et al.}, ``{Search for a Higgs Boson
  Decaying to Two $W$ Bosons at CDF},''
  \href{http://dx.doi.org/10.1103/PhysRevLett.102.021802}{{\em Phys. Rev.
  Lett.} {\bf 102} (2009)  021802},
\href{http://arxiv.org/abs/0809.3930}{{\tt arXiv:0809.3930 [hep-ex]}}.

\bibitem{D0WWnew}
{\bf D\O} Collaboration \href{http://arxiv.org/abs/Note 5757-CONF}{{\tt Note
  5757-CONF}}.

\bibitem{D0WW}
{\bf D\O} Collaboration \href{http://arxiv.org/abs/Note 5063-CONF}{{\tt Note
  5063-CONF}}.

\bibitem{Sjostrand:2006za}
T.~Sjostrand, S.~Mrenna, and P.~Skands, ``{PYTHIA 6.4 physics and manual},''
  {\em JHEP} {\bf 05} (2006)  026,
\href{http://arxiv.org/abs/hep-ph/0603175}{{\tt arXiv:hep-ph/0603175}}.

\bibitem{Asai:2004ws}
S.~Asai {\em et al.}, ``{Prospects for the search for a standard model Higgs
  boson in ATLAS using vector boson fusion},''
  \href{http://dx.doi.org/10.1140/epjcd/s2003-01-010-8}{{\em Eur. Phys. J.}
  {\bf C32S2} (2004)  19--54},
\href{http://arxiv.org/abs/hep-ph/0402254}{{\tt arXiv:hep-ph/0402254}}.

\bibitem{Aad:2009wy}
{\bf The ATLAS} Collaboration, G.~Aad {\em et al.}, ``{Expected Performance of
  the ATLAS Experiment - Detector, Trigger and Physics},''
\href{http://arxiv.org/abs/0901.0512}{{\tt arXiv:0901.0512 [Unknown]}}.

\bibitem{Yazgan:2007cd}
E.~Yazgan {\em et al.}, ``{Search for a standard model Higgs boson in CMS via
  vector boson fusion in the $H \to W W \to \ell \nu \ell \nu$ channel},''
  \href{http://dx.doi.org/10.1140/epjc/s10052-007-0485-2}{{\em Eur. Phys. J.}
  {\bf C53} (2008)  329--347},
\href{http://arxiv.org/abs/0706.1898}{{\tt arXiv:0706.1898 [hep-ex]}}.

\bibitem{Eboli:2000ze}
O.~J.~P. Eboli and D.~Zeppenfeld, ``{Observing an invisible Higgs boson},''
  \href{http://dx.doi.org/10.1016/S0370-2693(00)01213-2}{{\em Phys. Lett.} {\bf
  B495} (2000)  147--154},
\href{http://arxiv.org/abs/hep-ph/0009158}{{\tt arXiv:hep-ph/0009158}}.

\bibitem{Rainwater:1999sd}
D.~L. Rainwater and D.~Zeppenfeld, ``{Observing $H \to W^{(*)}W^{(*)} \to e^\pm
  \mu^\mp /\!\!\!{p}_T$ in weak boson fusion with dual forward jet tagging at
  the CERN LHC},'' \href{http://dx.doi.org/10.1103/PhysRevD.60.113004}{{\em
  Phys. Rev.} {\bf D60} (1999)  113004},
\href{http://arxiv.org/abs/hep-ph/9906218}{{\tt arXiv:hep-ph/9906218}}.

\bibitem{Kauer:2000hi}
N.~Kauer, T.~Plehn, D.~L. Rainwater, and D.~Zeppenfeld, ``{H --> W W as the
  discovery mode for a light Higgs boson},''
  \href{http://dx.doi.org/10.1016/S0370-2693(01)00211-8}{{\em Phys. Lett.} {\bf
  B503} (2001)  113--120},
\href{http://arxiv.org/abs/hep-ph/0012351}{{\tt arXiv:hep-ph/0012351}}.

\bibitem{Barr:2009mx}
A.~J. Barr, B.~Gripaios, and C.~G. Lester, ``{Measuring the Higgs boson mass in
  dileptonic W-boson decays at hadron colliders},''
\href{http://arxiv.org/abs/0902.4864}{{\tt arXiv:0902.4864 [hep-ph]}}.

\bibitem{Cho:2006sx}
G.~C. Cho {\em et al.}, ``{Weak boson fusion production of supersymmetric
  particles at the LHC},''
  \href{http://dx.doi.org/10.1103/PhysRevD.73.054002}{{\em Phys. Rev.} {\bf
  D73} (2006)  054002},
\href{http://arxiv.org/abs/hep-ph/0601063}{{\tt arXiv:hep-ph/0601063}}.

\bibitem{Sullivan:2008ki}
Z.~Sullivan and E.~L. Berger, ``{Trilepton production at the CERN LHC: Standard
  model sources and beyond},''
  \href{http://dx.doi.org/10.1103/PhysRevD.78.034030}{{\em Phys. Rev.} {\bf
  D78} (2008)  034030},
\href{http://arxiv.org/abs/0805.3720}{{\tt arXiv:0805.3720 [hep-ph]}}.

\bibitem{Aaltonen:2008mi}
{\bf CDF} Collaboration, T.~Aaltonen {\em et al.}, ``{Search for the Higgs
  boson in events with missing transverse energy and $b$ quark jets produced in
  proton- antiproton collisions at $\sqrt{s}$ =1.96 TeV},''
  \href{http://dx.doi.org/10.1103/PhysRevLett.100.211801}{{\em Phys. Rev.
  Lett.} {\bf 100} (2008)  211801},
\href{http://arxiv.org/abs/0802.0432}{{\tt arXiv:0802.0432 [hep-ex]}}.

\end{thebibliography}\endgroup
\bibliographystyle{utphys}

\end{document}